\documentclass[aps,prl,showpacs,twocolumn,superscriptaddress]{revtex4}

\usepackage{amsmath}
\usepackage{amssymb}
\usepackage{bm}
\usepackage{natbib}
\usepackage{graphicx}
\usepackage{color}
\usepackage{dcolumn}

\begin{document}


\title{Band Gap and Edge Engineering via Ferroic Distortion and Anisotropic Strain: The Case of SrTiO$_{3}$}

\author{Robert F. Berger}
\affiliation{Molecular Foundry, Lawrence Berkeley National Laboratory, Berkeley, CA}
\author{Craig J. Fennie}
\affiliation{School of Applied and Engineering Physics, Cornell University, Ithaca, NY}
\author{Jeffrey B. Neaton}
\email{jbneaton@lbl.gov}
\affiliation{Molecular Foundry, Lawrence Berkeley National Laboratory, Berkeley, CA}


\begin{abstract}
The effects of ferroic distortion and biaxial strain on the band gap and band edges of SrTiO$_{3}$ (STO) are calculated using density functional theory and many-body perturbation theory. Anisotropic strains are shown to reduce the gap by breaking degeneracies at the band edges. Ferroic distortions are shown to widen the gap by allowing new band edge orbital mixings. Compressive biaxial strains raise band edge energies, while tensile strains lower them. To reduce the STO gap, one must lower the symmetry from cubic while suppressing ferroic distortions. Our calculations indicate that for engineered orientation of the growth direction along $[111]$, the STO gap can be controllably and considerably reduced at room temperature.
\end{abstract}

\date{\today}


\maketitle


Within the perovskite family (ABO$_{3}$), a remarkable range of functional phenomena---ferroelectricity, ferromagnetism, multiferroicity, and superconductivity, for example---can be accessed through chemical substitution at the A and B sites.~\cite{wolfram06} In recent years, biaxial strains induced by epitaxy have provided another route to tune perovskite properties.~\cite{dawber05rmp} In a well-known example, both theory and experiment have shown that thin films of SrTiO$_{3}$ (STO), a prototypical perovskite insulator that takes up a centrosymmetric cubic structure ($Pm\overline{3}m$) under standard conditions, become polar and ferroelectric at room temperature for sufficiently large biaxial strains induced by coherent epitaxial growth.~\cite{pertsev00prb,haeni04n} In STO, the large biaxial strains achievable via epitaxy provide access to a rich landscape of ferroic distortions, both polar ferroelectric and nonpolar antiferrodistortive,~\cite{pertsev00prb,antons05prb,dieguez05prb} that are known to alter both ferroelectric polarization and the low-frequency dielectric response.~\cite{antons05prb}

In addition to its well-studied dielectric response, STO has also long been explored in solar water splitting applications due to its favorable flat-band potentials and stability against degradation in water.~\cite{bak02ijhe} However, its indirect optical gap (3.2 eV~\cite{cardona65pr}) is too large for efficient light harvesting in the solar spectrum. Prior work has focused on chemical substitution as a means for reducing perovskite gaps,~\cite{fujimori92prb} but with unintended consequences for other properties. In this Letter, we use density functional theory (DFT) and beyond to examine a new route, via anisotropic strain and ferroic distortions, for altering the band gap and band edges of STO. We show that while anisotropic strain in the absence of ferroic distortions reduces the gap, ferroic distortions increase it. In addition, band edge energies, important for electro- and photocatalysis, can also be controllably increased or reduced relative to vacuum. The predicted trends can be fully understood with symmetry arguments, suggesting well-defined routes for tuning the electronic structures of perovskite oxides. For example, using intuition developed here, we show that strained thin films of STO grown along an alternate orientation, $[111]$, have considerably reduced gaps relative to bulk at room temperature.

STO structural and electronic properties are obtained using DFT within the local density approximation (LDA), using the VASP package~\cite{kresse96prb} and PAW potentials.~\cite{kresse99prb} Electrons taken to be valence are $4s^{2}4p^{6}5s^{2}$ of Sr, $3s^{2}3p^{6}4s^{2}3d^{2}$ of Ti, and $2s^{2}2p^{4}$ of O. A plane-wave cutoff of 500 eV is used throughout. At room temperature, STO takes up the cubic perovskite structure ($Pm\overline{3}m$) with a five-atom primitive cell, which we treat with a $\Gamma$-centered $8\times8\times8$ $k$-point mesh. Near 100 K, STO is known to undergo a transition to an antiferrodistortive phase in which Ti--O octahedra rotate, doubling the unit cell and lowering symmetry to $I4/mcm$.~\cite{fleury68prl} For such phases requiring supercells, we use proportionally fewer $k$-points. These parameters result in excellent convergence and good agreement with previous work.

\begin{figure}
  \begin{center}
    \includegraphics[scale=0.2]{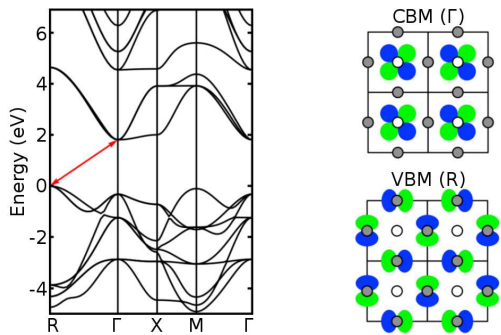}
  \end{center}
  \caption{(Left) DFT-LDA band structure of cubic STO. (Right) Schematic pictures of the highest occupied and lowest unoccupied states of STO as linear combinations of atomic orbitals.  Ti atoms are shown in white, O atoms in gray.}
  \label{sto}
\end{figure}

The DFT-LDA near-gap band structure of optimized cubic STO ($a$=3.864 {\AA}) is shown in Fig.~\ref{sto}. The conduction band minimum (CBM) lies at $\Gamma$ and consists almost entirely of Ti $3d$ states, while the valence band maximum (VBM) lies at R and consists almost entirely of O $2p$ states. As expected within DFT-LDA, the calculated indirect gap of 1.80 eV is far narrower than the experimental optical gap of 3.2 eV.~\cite{cardona65pr} To obtain reliable band gaps and edges, we use many-body perturbation theory within the GW approximation~\cite{shishkin06prb}, known to give quantitative band gaps and band edge energies for select semiconducting systems in the absence of strong excitonic effects.~\cite{hybertsen86prb} Our computed G$_{0}$W$_{0}$ gap of cubic STO (see~\footnote{G$_{0}$W$_{0}$ corrections to DFT-LDA eigenvalues are computed in VASP using a frequency-dependent RPA dielectric function sampling 96 frequency points, 288 bands (268 unfilled), a plane-wave cutoff for the response function of 300 eV, and a $\Gamma$-centered $6\times6\times6$ $k$-point mesh.}) is 3.82 eV, 0.6 eV larger than experiment.~\footnote{In our STO calculations, 0.1 eV of the difference between G$_{0}$W$_{0}$ and experimental gaps is due to our use of the DFT-LDA optimized lattice parameter, which is more than 1\% shorter than that of experiment.} This overestimate is consistent with past calculations for STO~\cite{cappellini00jpcm} and a closely-related system, TiO$_{2}$, where deviations are suggested to arise from excitonic and polaronic effects.~\cite{kang10prb} Overestimate notwithstanding, our G$_{0}$W$_{0}$ band edge energies at equilibrium are in reasonable agreement with experiment. Experiments place the CBM and VBM 2.6 and 5.7 eV below the vacuum level, respectively,~\cite{maus02ss} compared with our G$_{0}$W$_{0}$ calculated values of 2.48 and 6.30 eV (see~\footnote{Band edge energies are calculated in two steps. A slab of six perovskite layers with unrelaxed TiO$_{2}$-terminated $(001)$ surfaces and 15 {\AA} of vacuum is used to reference a bulk-like Ti $3s$ semicore state to the vacuum level. A periodic calculation is then used to compute band edge energies relative to the same semicore state.}).

Following previous calculations,~\cite{antons05prb,dieguez05prb} we model biaxially strained thin films of STO by periodic calculations in which in-plane lattice parameters are fixed while the perpendicular axis and atomic positions are allowed to relax. Our structures have in-plane strains of up to $\pm4\%$, and the resulting lattice parameters (Supplemental Information) and structural phase diagram agree well with previous work.~\cite{pertsev00prb,antons05prb,dieguez05prb} This range of strain is somewhat optimistic experimentally, as strains approaching $\pm4\%$ could limit the achievable film thickness, making bulk optical measurements difficult, and potentially introducing interfaces and quantum confinement effects. Nonetheless, perovskite substrates that would induce these strains are available,~\cite{schlom07armr} and there are no fundamental limitations on stabilization of thin films on these substrates. For ultrathin films, our results here represent a quantitative baseline upon which future studies of quantum confinement and interface effects may be understood.

\begin{figure*}
  \begin{center}
    \includegraphics[scale=0.2]{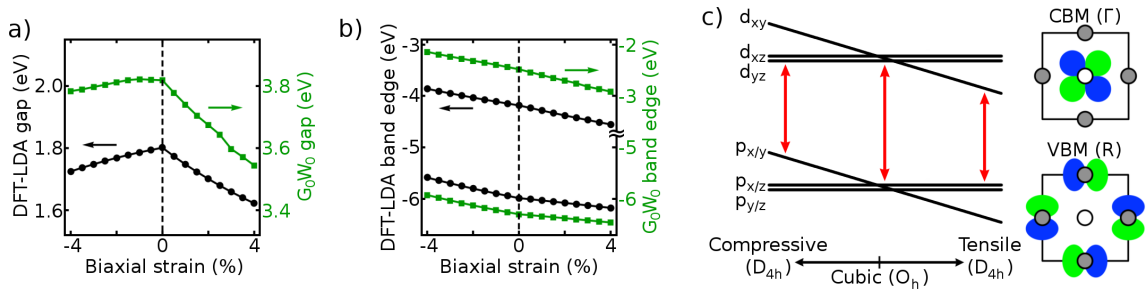}
  \end{center}
  \caption{a) Computed gaps and b) band edges of paraelectric, biaxially reshaped STO ($P4/mmm$ symmetry) under biaxial strain perpendicular to $[001]$.  c) Schematic illustration of the trends in band edges as a function of strain.}
  \label{reshaping}
\end{figure*}

We start with the most common growth orientation, [001], and compute electronic gaps and band edge energies for biaxial strains up to $\pm4\%$; these are shown in Fig.~\ref{reshaping}, initially for the paraelectric case in which the $[001]$ axis is relaxed but further distortion of the internal coordinates is forbidden ($P4/mmm$ symmetry). Interestingly, biaxial strains tune both CBM and VBM by several tenths of an eV, suggesting a route to optimize each band edge relative to the potentials of the water splitting half-reactions. From Fig.~\ref{reshaping}b, compressive biaxial strains raise band edge energies relative to vacuum, and tensile biaxial strains lower these energies. For paraelectric STO, the DFT-LDA gap narrows under both compressive and tensile strain. That the gap is non-monotonic about the cubic structure is expected on symmetry grounds (Fig.~\ref{reshaping}c). Cubic STO has $O_{h}$ point group symmetry, and both its CBM and VBM are threefold-degenerate. Biaxial reshaping of the unit cell lowers the point group symmetry to $D_{4h}$, splitting these degenerate triples into a spread of singles and pairs. As a result, the gap is narrower for tetragonal paraelectric STO than for cubic STO.

Notably, our G$_{0}$W$_{0}$ corrections to Kohn-Sham DFT-LDA eigenvalues (Fig.~\ref{reshaping}a,b, green) remain quite constant with changing biaxial strain. (A similar result was reported for silicon under isotropic pressure.~\cite{zhu89prb}) The same is true of G$_{0}$W$_{0}$ corrections to STO structures with ferroic distortions (Supplemental Information). This supports the premise that DFT-LDA, while insufficient for producing numerically accurate optical gaps, correctly tracks their {\em trends}.~\cite{zhu89prb,zhao04prl} For the remainder of this paper, we restrict our focus to DFT-LDA results, at times assuming a constant correction of 1.4 eV---the difference between DFT-LDA and room-temperature optical gaps of equilibrium STO---to provide concrete estimates for gaps under biaxial strain.

We turn now to the issue of how ferroic distortions affect the STO band gap. Two classes of commonly studied ferroic distortions in perovskites are ferroelectric polar modes (FE, relative translation of the cation and anion sublattices) and antiferrodistortive modes (AFD, collective rotation of the Ti--O octahedra). While neither distortion is present in unstrained STO at room temperature, both are accessible via strain, temperature, and applied fields.~\cite{fleury68prl,pertsev00prb,haeni04n} To illustrate how these distortions might affect the band gap, we freeze in displacement patterns associated with the eigenvectors of low-lying IR-active phonon modes along the $[001]$, $[110]$, and $[111]$ axes, and examine the DFT-LDA gap as a function of amplitude. We freeze in AFD modes by interpolating between cubic STO and optimized structures with AFD rotations about the $[001]$, $[110]$, and $[111]$ axes.

\begin{figure}
  \begin{center}
    \includegraphics[scale=0.2]{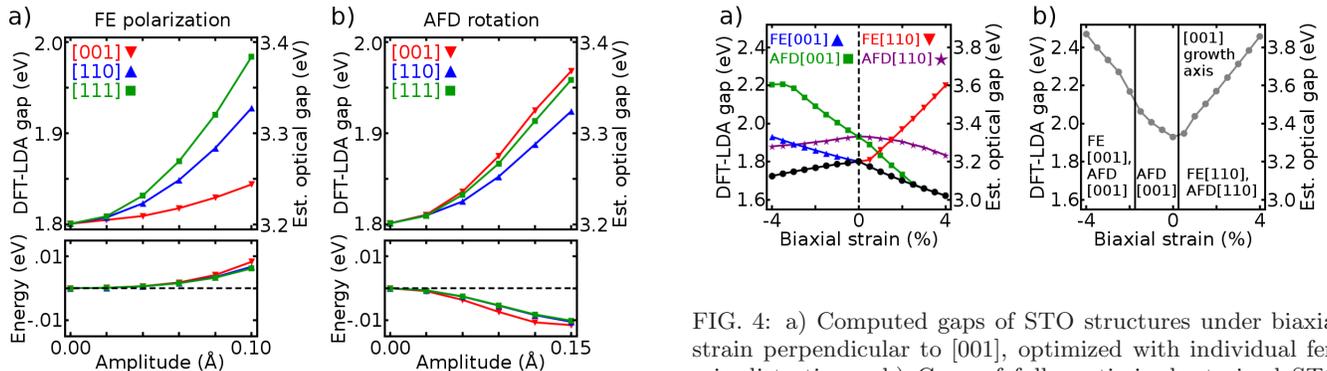}
  \end{center}
  \caption{Computed gaps (top) of STO structures in which a) FE polarization modes and b) AFD rotation modes are frozen in. Per-formula-unit structural energies relative to cubic STO are also shown (bottom). FE amplitudes are expressed as average relative translations of Ti and O atoms, while AFD amplitudes are expressed as average translations of O atoms. AFD rotations refer to $[001]=a^{0}a^{0}c^{-}$, $[110]=a^{-}a^{-}c^{0}$, and $[111]=a^{-}a^{-}a^{-}$ in Glazer notation.~\cite{glazer72ac}}
  \label{frozengaps}
\end{figure}

Computed gaps of these distorted STO structures are shown in Fig.~\ref{frozengaps}. For modest changes in structural energy, both FE (Fig.~\ref{frozengaps}a) and AFD (Fig.~\ref{frozengaps}b) modes widen the gap by as much as 0.2 eV. This widening can be understood with symmetry arguments. As the symmetry is lowered under both types of distortion, conduction (primarily Ti $3d$) and valence (primarily O $2p$) states are allowed to mix and repel one another. Table~\ref{symtable} shows the irreducible representations of Ti $3d$ and O $2p$ states at $\Gamma$ and R within the $O_{h}$ point group. By symmetry, the CBM of cubic STO cannot mix with O $2p$ states at $\Gamma$ or R. In the presence of FE distortions, inversion symmetry is broken, the distinction between even and odd ($g$ and $u$) parity is eliminated, and the CBM is allowed to mix with O $2p$ states. This mixing was previously illustrated by Wheeler et al.~\cite{wheeler86jacs} for STO in terms of linear combinations of atomic orbitals. Similarly, in the presence of AFD rotations, doubling of the unit cell renders $\Gamma$ and R equivalent, allowing the CBM to mix with O $2p$ states.  Likewise, the VBM of STO can mix with Ti $3d$ states only in the presence of FE and/or AFD distortions.  As Ti $3d$ and O $2p$ states mix, they widen the STO band gap.  While the magnitude of the widening depends on the geometry of a given distortion, this trend is general.  Interestingly, AFD rotations can transform the STO gap from indirect to direct, as has been previously noted in both theory~\cite{galinetto06f,heifets06jpcm} and experiment.~\cite{galinetto06f,shablaev84zetf}

\begin{table}
  \caption{Irreducible representations of Ti $3d$ and O $2p$ states at $\Gamma$ and R within $O_{h}$ point group symmetry.  The CBM and VBM of cubic STO are shown in red.}
  \centering
  \begin{tabular}{l@{\hspace{3ex}}l}
  \hline
  Atomic orbitals & Irreducible representations\\
  \hline
  Ti $3d$ at $\Gamma$ & {\color{red}{$t_{2g}$}}, $e_{g}$\\
  Ti $3d$ at R & $t_{2g}$, $e_{g}$\\
  O $2p$ at $\Gamma$ & $t_{1u}$, $t_{1u}$, $t_{2u}$\\
  O $2p$ at R & {\color{red}{$t_{1g}$}}, $t_{2g}$, $e_{g}$, $a_{1g}$\\
  \hline
  \end{tabular}
  \label{symtable}
\end{table}

\begin{figure}
  \begin{center}
    \includegraphics[scale=0.2]{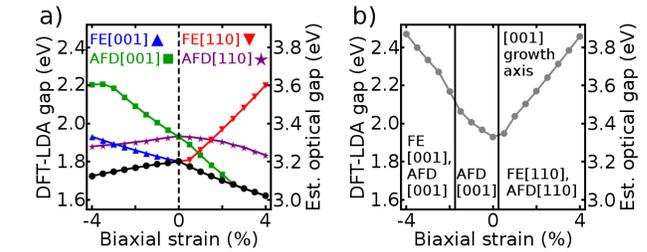}
  \end{center}
  \caption{a) Computed gaps of STO structures under biaxial strain perpendicular to $[001]$, optimized with individual ferroic distortions.  b) Gaps of fully optimized, strained STO structures, computed at zero temperature.}
  \label{zero}
\end{figure}

Having established how anisotropic strain and ferroic distortions impact the STO electronic structure separately, we present in Fig.~\ref{zero} the computed gaps of structures with ferroic distortions under biaxial strain (assuming a $[001]$ growth direction). The energetic landscape of ferroic distortions in ground-state STO has been studied extensively within DFT-LDA.~\cite{zhong95prl,antons05prb,dieguez05prb} Consistent with this work, full relaxation causes FE and AFD to combine along $[001]$ under compressive strain, and along $[110]$ under tensile strain. All ferroic distortions, optimized separately (Fig.~\ref{zero}a), widen the gap relative to paraelectric STO. Full atomic relaxation (Fig.~\ref{zero}b) widens the gap even further. Using biaxial strain and associated ferroic distortions, we predict the optical gap can be tuned by nearly 20\% under $\pm4$\% strains.

As we have shown, the breaking of symmetry in STO has two manifestations with competing effects on the band gap---biaxial reshaping of the unit cell that narrows the gap through broken degeneracies, and ferroic distortions that widen the gap through new orbital mixings. The data in Fig.~\ref{zero}b suggest that, in biaxially strained STO at low temperatures, ferroic distortions have the dominant effect on the gap. To engineer a {\it narrower} STO band gap, we must reshape the unit cell while suppressing ferroic distortions. One way to suppress ferroic distortions is to work at room temperature. Using a Landau-Ginzburg-Devonshire model with empirical parameters, Pertsev et al.~computed a phase diagram of STO under $[001]$ biaxial strain.~\cite{pertsev00prb} At 300 K, their model predicts that STO is paraelectric between 1.8\% compressive and 1.5\% tensile strain. While this provides a range of strain within which the STO gap can be narrowed, ferroic distortions widen the gap outside the range.

\begin{figure}
  \begin{center}
    \includegraphics[scale=0.2]{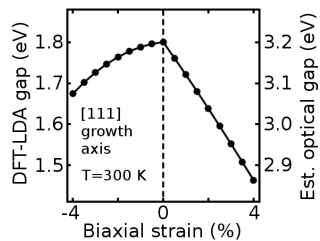}
  \end{center}
  \caption{Computed gaps of STO using free energy minimized structures at 300 K under biaxial strain perpendicular to $[111]$.  All structures are paraelectric.}
  \label{room}
\end{figure}

If, however, STO is grown epitaxially along an axis other than $[001]$, the possibility exists for reshaping the unit cell without inducing ferroic distortions. Unlike growth along $[001]$, which forces in-plane Ti--O distances to be too short or too long for optimal bonding, a growth axis of $[111]$ imposes no constraints on Ti--O distances. We therefore expect that the $[111]$ case does not strongly enhance ferroic distortions in STO. Using the same empirical parameters of Ref.~\cite{pertsev00prb}, but applying them to films grown along $[111]$ at 300 K, confirms that STO is expected to remain paraelectric over the entire range within $\pm4\%$ strain.  We therefore predict that $[111]$ strain at room temperature (achievable via epitaxial growth on the $(111)$ surfaces of perovskite substrates) could narrow the STO band gap significantly more than $[001]$ strain (Fig.~\ref{room}).  Within DFT-LDA, 4\% tensile strain perpendicular to $[111]$ narrows the Kohn-Sham gap by more than 0.3 eV, to 1.46 eV.  Using our constant correction, we therefore estimate that this approach could lower the optical gap of STO below 3 eV.

To conclude, we have examined the extent to which engineering of the STO crystal structure can tune its electronic gap and band edges. While biaxial strain widens the STO gap at zero temperature via ferroic distortions, it has the ability to narrow the gap significantly at room temperature, especially when applied perpendicular to the $[111]$ direction.  The observed trends can be understood on general symmetry grounds, suggesting that similar intuition can be productively applied across the broad family of perovskites.

We are indebted to D.~G. Schlom for inspiring this project and insightful discussions. This material is based upon work supported as part of the Energy Materials Center at Cornell (EMC$^{2}$), an Energy Frontier Research Center funded by the U.S.~Department of Energy, Office of Science, Office of Basic Energy Sciences under Award Number DE-SC0001086. Work at the Molecular Foundry was supported by the Office of Science, Office of Basic Energy Sciences, of the U.S.~Department of Energy under Contract Number DE-AC02-05CH11231.

\bibliographystyle{prsty}
\bibliography{sto.bib}


\end{document}